\documentclass[sigconf]{acmart}

\AtBeginDocument{%
  }

\setcopyright{acmlicensed}

\copyrightyear{2026}
\acmYear{2026}
\setcopyright{cc}
\setcctype{by}
\acmConference[MM '26]{Proceedings of the 34th ACM International Conference on Multimedia}{November 10--14, 2026}{Rio de Janeiro, Brazil}
\acmBooktitle{Proceedings of the 34th ACM International Conference on Multimedia (MM '26), November 10--14, 2026, Rio de Janeiro, Brazil}
\acmDOI{10.1145/3767308.3834925}
\acmISBN{979-8-4007-2213-4/2026/11}

\usepackage{multirow}
\usepackage{multicol}
\usepackage{algorithm}
\usepackage{algorithmic}

\usepackage{subcaption}

\acmSubmissionID{171}

\begin{document}

\title{Mind the Couch! Eliciting MLLM Reasoning in Interior Design via Weak-to-Strong Task Vector Injection}

\author{Yuxuan Yang}
\orcid{0009-0008-7180-1397}
\authornote{These authors have contributed equally}
\affiliation{%
  \institution{Nanjing Forestry University}
  \city{Nanjing}
  \country{China}
}
\email{qingzhuo13@njfu.edu.cn}

\author{Jingyao Wang}
\orcid{0000-0003-1782-8704}
\authornotemark[1]
\authornote{corresponding Author}
\affiliation{%
  \institution{Institute of Software Chinese Academy of Sciences}
  \institution{University of the Chinese Academy of Sciences}
  \city{Beijing}
  \country{China}
}
\email{wangjingyao2023@iscas.ac.cn}

\author{Luntian Mou}
\orcid{0000-0002-1551-4448}
\affiliation{%
  \institution{Beijing University of Technology}
  \institution{Beijing Key Laboratory of Embodied Interactive Intelligence}
  \city{Beijing}
  \country{China}
}
\email{ltmou@pku.edu.cn}



\begin{abstract}
  Multimodal Large Language Models (MLLMs) have demonstrated great performance, yet they often suffer from severe modality misalignment when confronted with densely constrained spaces for interior design. Due to the loss of high-frequency local topological details and fine-grained aesthetic shifts during visual encoding, existing MLLMs frequently hallucinate, yielding physical spatial collisions and visual aesthetic dissonance. To address this, we propose \textbf{D}ual-prior \textbf{A}ctivation \textbf{R}esidual \textbf{T}ask-vectors \textbf{I}njection mechanism (DART-I) for MLLMs. It shifts the paradigm from lossy text-prompting to direct latent intervention, utilizing weak-to-strong deterministic rules to anchor the causal reasoning of MLLMs for interior design. Specifically, DART-I operates in three steps: it first explicitly extracts continuous spatial distance and color typography features from images using extremely lightweight weak experts; subsequently, it transforms these deterministic priors into directional task vectors via a linear projection network; these vectors are dynamically injected as residual terms into the latent space of the frozen MLLMs, steering MLLMs towards precise reasoning for interior design. Stepping outside the conventional paradigms, our method achieves precise reasoning without fine-tuning the MLLMs, effectively bypassing expensive computational costs and catastrophic forgetting. Extensive experiments on various benchmarks demonstrate the effectiveness and advantages of DART-I.
\end{abstract}

\ccsdesc[500]{Applied computing~Computer-aided design}
\ccsdesc[500]{Computing methodologies~Machine learning}
\ccsdesc[500]{Computing methodologies~Spatial and physical reasoning}
\keywords{interior design; multimodal large language model; weak-to-strong task vector injection}



\maketitle

\section{Introduction}
\label{sec:intro}
\begin{figure}[t]
    \centering
    \includegraphics[width=\linewidth]{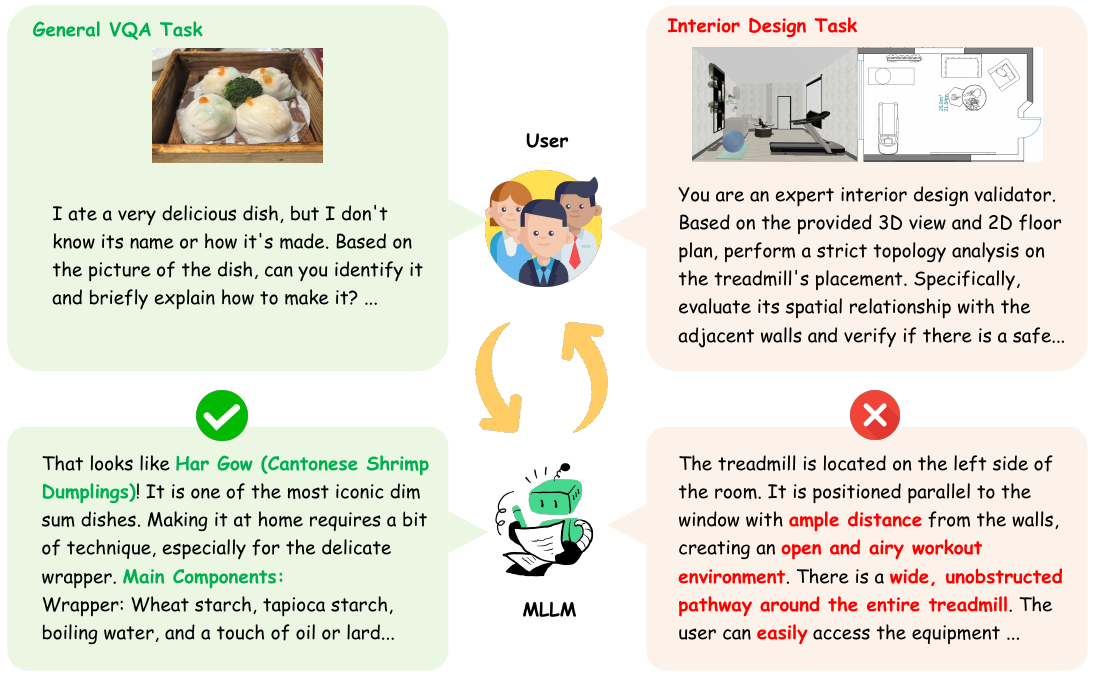}
    \vspace{-0.2in}
    \caption{An example of MLLM on a general VQA task (left) versus an interior design evaluation task (right).}
    \label{fig:intro}
    \vspace{-0.15in}
\end{figure}

Multimodal Large Language Models (MLLMs) have demonstrated remarkable capabilities in various tasks, e.g., scene understanding and visual question-answering tasks \cite{liu2023visual,ccelen2024design,gallega2025exploring,guo2025copo}. However, when these general-purpose models are transferred to highly specialized scenarios, such as the field of interior design \cite{,pile2005history,ulrich1991effects}, their reasoning performance often faces generalization challenge (\textbf{Figure \ref{fig:intro}}). Unlike natural wide-area scenes, which feature relatively sparse feature distributions and high error tolerance, interior design is fundamentally a densely constrained restricted space  \cite{torres2017fault,ching2018interior,wang2024root}. It requires not only strict physical geometric arrangements (e.g., precise layout topologies and collision-free circulation paths) but is also governed by rigorous visual aesthetic principles (e.g., harmony in color contrast and balance of visual weight). This leap from macroscopic loose semantic perception to dense rule reasoning makes it difficult for existing MLLMs to output reliable conclusions that comply with both engineering and aesthetic standards when processing professional design blueprints and renderings.

\begin{figure*}[t]
    \centering
    \includegraphics[width=\linewidth]{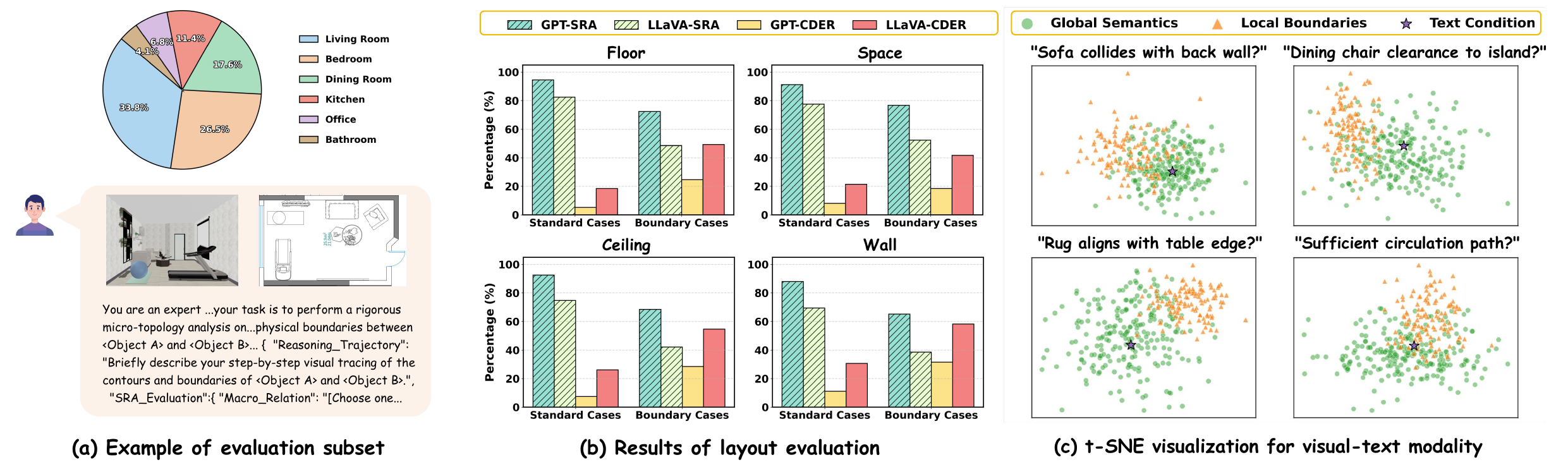}
    \caption{Motivation and performance analysis of MLLMs in interior design tasks.
(a) shows category distribution of the evaluation dataset and an example of the structured prompt used for topology probing. 
(b) provides the performance of different models when transitioning to boundary cases across four spatial dimensions. 
(c) shows the t-SNE visualization results.}
    \label{fig:motivation}
\end{figure*}

To investigate the origins of the aforementioned generalization bottleneck, we conduct a series of experiments for analyzing MLLMs in interior design tasks (\textbf{Subsection \ref{sec:empirical_evidence}}). We collect examples from Structured3D \cite{zheng2020structured3d} and 3D-FRONT \cite{fu20213d} (e.g., topological boundary cases with extremely narrow furniture spacing, and spaces with extreme color contrasts, as shown in \textbf{Figure \ref{fig:motivation}(a)}). These are fed into MLLMs (including the closed-source GPT-4V and the open-source LLaVA-v1.6-7B) for zero-shot spatial relation and aesthetic question answering. The quantitative observations (\textbf{Figure \ref{fig:motivation}(b)}) indicate that when confronted with fine-grained physical boundaries, the spatial relation error rates of all baseline models increase significantly.
To deconstruct this failure mechanism, we further visualize the feature responses within the cross-modal attention layers of LLaVA-v1.6-7B. As shown in \textbf{Figure \ref{fig:motivation}(c)}, we observe a pronounced modality misalignment: when processing high-frequency constraint instructions such as ``collision'' or ``spacing'', the model's visual attention weights remain highly concentrated on the global semantic centers of objects (e.g., the geometric centers of sofas or beds). Conversely, its activation responses to the subtle boundary contours that govern physical topologies are extremely sparse or even entirely absent. This indicates that the model loses high-frequency local details during feature extraction, depriving the autoregressive decoder of precise visual conditioning. Consequently, this leads to frequent spatial hallucinations and aesthetic misjudgments during the text generation stage.

To address the inadequate perceptual granularity of MLLMs, some concurrent methods \cite{chen2025mindgpt,yang2025optiscene,mao2025spatiallm} typically rely on fine-tuning using massive blueprint-expert review datasets. However, acquiring high-quality, fully annotated multimodal data in the architectural design domain is prohibitively expensive \cite{dai2017scannet,wang2023amsa}. More critically, deep domain-specific fine-tuning easily disrupts the original parameter distribution of MLLMs, leading to catastrophic forgetting of general commonsense \cite{luo2025empirical,wang2025learning,zhu2024model}. Therefore, we abandon the costly fine-tuning route and propose a weak-to-strong representation intervention approach to anchor the model's underlying causal reasoning process. This paradigm perfectly aligns with the core requirements of interior design tasks: comprehensively evaluating design schemes demands not only high-level semantic and aesthetic cognition (i.e., the strength of strong models) but also strict adherence to physical boundaries and topological constraints (i.e., the blind spot of strong models). Although MLLMs tend to lose high-frequency local details during visual encoding, extremely lightweight traditional vision algorithms or non-parametric models (i.e., weak experts) possess high mathematical certainty and sensitivity to underlying geometric contours, pixel-level spacing, and fundamental color distributions \cite{voulodimos2018deep,stricker1995similarity}. Based on this, we propose introducing lightweight weak experts to explicitly extract continuous spatial topology and aesthetic layout features to construct task vectors, which are then embedded into MLLMs to guide precise reasoning for interior design without training costs.

Building upon these insights, we propose a novel Dual-prior Activation Residual Task-vectors Injection (DART-I) mechanism for MLLMs. Its core lies in shifting the paradigm from lossy fine-tuning and text-prompting to direct latent injection, achieving precise spatial and aesthetic reasoning of MLLMs with limited costs. Specifically, DART-I first utilizes extremely lightweight weak experts to explicitly extract continuous spatial distance and color layout features from images. Subsequently, it transforms these deterministic physical priors into directional task vectors via a projection network. By dynamically injecting these vectors as residual terms into the latent space of the frozen MLLMs, DART-I steers their reasoning process for interior design. 
Extensive experiments on various benchmarks, comprising CAD floor plans and high-definition renderings, demonstrate the advantage of DART-I. 
In summary, the main contributions can be summarized as:
\begin{itemize}
    \item We reveal an important issue in MLLMs for interior design via empirical analyses: the loss of high-frequency topological details deprives models of precise visual grounding, causing spatial hallucinations and aesthetic misjudgments.
    \item We propose DART-I, a novel weak-to-strong Dual-prior Activation Residual Task-vectors Injection mechanism for MLLMs. By extracting spatial and aesthetic priors via lightweight weak experts and dynamicly injecting them as residual task vectors into the latent space, it advances causal reasoning for interior design with limited costs.
    \item Across diverse interior design benchmarks, DART-I achieves superior performance. It significantly reduces spatial collision errors and aligns aesthetic evaluations with human experts at minimal computational cost.
\end{itemize}

\section{Related Work}
\label{sec:related_work}

\paragraph{\textbf{Interior Design}}
Interior design is fundamentally a densely constrained problem that requires strict adherence to both precise geometric topologies (e.g., collision-free circulation paths) and rigorous visual aesthetic principles \cite{ching2018interior,pile2005history,ulrich1991effects}. While early computational approaches and recent data-driven pipelines \cite{ching2018interior,nauata2020house,hu2020graph2plan,li2019layoutgan,wang2019planit}, such as graph-constrained layout synthesis and diffusion-based rendering, have advanced the field, they typically optimize for either structural rules or perceptual realism in isolation. Consequently, structurally valid layouts often lack aesthetic nuance, whereas visually stunning generative renderings frequently hallucinate physically impossible geometries. To address this leap from macroscopic loose semantic perception to microscopic dense rule reasoning, recent solutions \cite{chen2025mindgpt,yang2025optiscene,mao2025spatiallm} often resort to deep domain-specific fine-tuning, which incurs prohibitively high data collection costs and easily disrupts the original parameter distribution of foundation models. Breaking away from these expensive paradigms, our work approaches computational interior design through a novel weak-to-strong representation intervention framework, explicitly extracting continuous spatial and aesthetic priors to satisfy both engineering and visual standards without requiring extensive model retraining.

\paragraph{\textbf{Multimodal Large Language Models}}
MLLMs typically couples a visual encoder with a large language model (LLM) to enable unified reasoning over images and text, which have shown strong performance across a wide range of tasks \cite{alayrac2022flamingo,li2023blip,liu2023visual,ouyang2022training}. However, in practical systems, their zero-shot reasoning can degrade substantially in highly specialized domains \cite{bai2024hallucination,jiang2024hallucination,wang2024mllm,huang2024visual,fei2024fine}, such as interior design where decisions hinge on fine-grained geometric feasibility and subtle stylistic cues. 
Our empirical analyses reveal that when processing densely constrained restricted spaces, existing MLLMs suffer from severe modality misalignment: they tend to over-focus on the global semantic centers of objects, systematically losing the high-frequency topological details and subtle boundary contours that govern physical topologies. This loss deprives the precise visual grounding, leading to frequent spatial hallucinations and aesthetic misjudgments. To overcome these limitations, we shift the alignment paradigm to weak-to-strong latent intervention. By transforming physical and aesthetic priors into directional task vectors and dynamically injecting them as residual terms into the latent space of frozen MLLMs, we effectively anchor the model's causal reasoning, enabling precise spatial evaluation.

\section{Problem Analysis and Motivation}
\label{sec:problem_analysis}
In this section, we begin by formulating the problem settings and specific constraints of interior design tasks (\textbf{Subsection \ref{sec:problem_settings}}). Next, we provide empirical evidence illustrating the modality misalignment and spatial hallucination bottlenecks in existing MLLMs when confronted with highly constrained spaces (\textbf{Subsection \ref{sec:empirical_evidence}}).

\subsection{Problem Settings}
\label{sec:problem_settings}

Consider a multimodal input sequence $I = (I_v, I_t)$, where $I_v$ represents the visual input (e.g., interior design blueprints or renderings) and $I_t$ denotes the textual instruction. A standard MLLM first encodes $I_v$ into a sequence of visual tokens via a vision encoder, which is then concatenated with the tokenized text $I_t$. The language model processes this joint sequence to autoregressively generate the output text $y=\{y_1, y_2, \ldots, y_T\}$:
\begin{equation}
p_\theta(y \mid I) = \prod_{t=1}^{T} p_\theta(y_t \mid I, y_{<t}),
\end{equation}
where $\theta$ represents the frozen parameters of the MLLM, and $T$ is the sequence length.

To bridge the gap between macroscopic generation and our latent intervention mechanism, we must formulate the internal forward pass. At each decoding step $t$, the transformer layers map the input tokens into a sequence of continuous latent representations. Let $h_t^{(l)} \in \mathbb{R}^d$ denote the hidden state at the $l$-th transformer layer. The final next-token distribution is predicted by a language modeling head over the last layer's representation $h_t^{(L)}$:
\begin{equation}
p_\theta(y_t \mid I, y_{<t}) = \text{Softmax}(W_{U} h_t^{(L)}),
\end{equation}
where $W_{U}$ is the unembedding matrix. Standard zero-shot inference directly samples from this distribution.

Unlike general tasks, interior design acts as a densely structured reasoning problem. For a given architectural input $I_v$, the generated output $y$ (which can be a spatial layout evaluation, a placement proposal, or a rendering description) must satisfy a rigid set of micro-constraints $\mathcal{C} = \mathcal{C}_{spa} \cup \mathcal{C}_{aes}$. In other words, an ideal MLLM policy should concentrate its probability mass strictly within the feasible region defined by these dual constraints:
\begin{equation}
\forall y \sim p_\theta(\cdot \mid I), \quad y \models (\mathcal{C}_{spa} \land \mathcal{C}_{aes}).
\end{equation}
where $\mathcal{C}_{spa}$ defines the strict physical and topological rules (e.g., minimum collision-free distance, feasible circulation paths, and absolute boundary limitations); $\mathcal{C}_{aes}$ encapsulates the visual aesthetic principles (e.g., color contrast harmony and typographical balance).

\begin{figure*}[t]
    \centering
    \includegraphics[width=\linewidth]{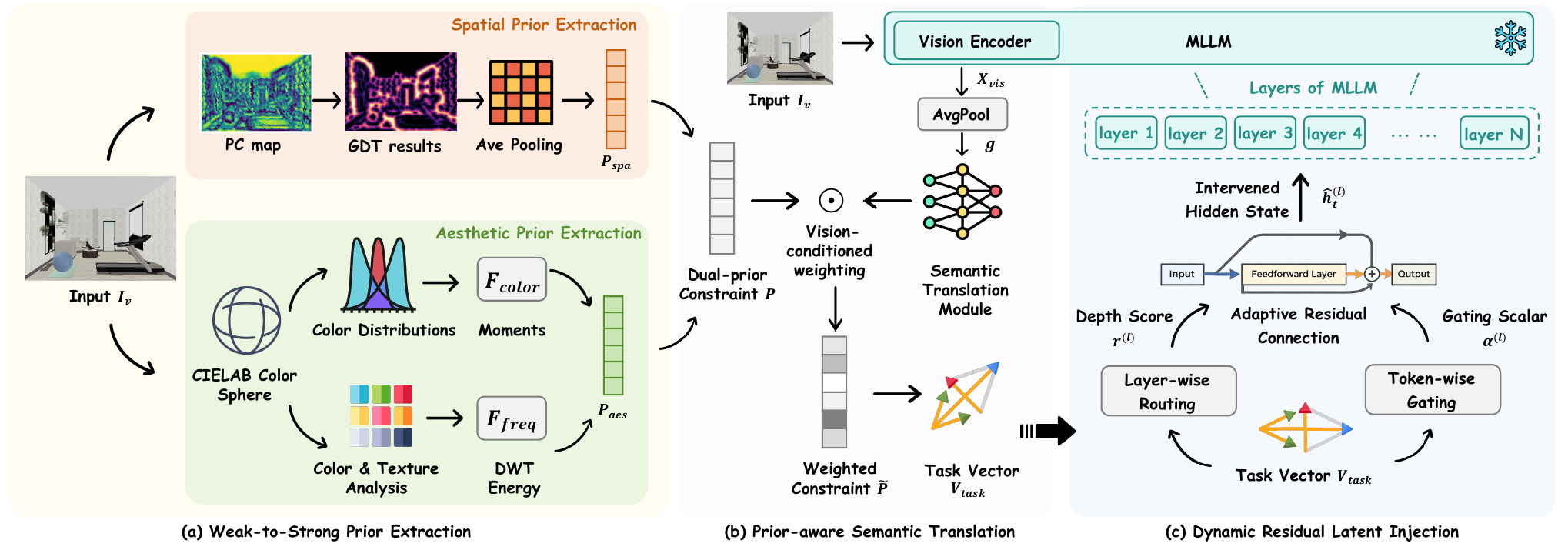}
    \vspace{-0.2in}
    \caption{Overview of DART-I. It first performs (a) weak-to-strong prior extraction to obtain spatial and aesthetic priors from the input. These priors undergo (b) prior-aware semantic translation to generate a task vector $V_{task}$. Finally, (c) dynamic residual latent injection applies this vector into the MLLM layers for interior design. The pseudo-code is shown in Appendix B.}
    \label{fig:framework}
\end{figure*}

\subsection{Empirical Evidence}
\label{sec:empirical_evidence}
To investigate whether existing MLLMs can reliably evaluate designs under the dense constraints of interior spaces, in this subsection, we explore the limits of their generalization capabilities in interior design tasks. Unlike general open-domain scene understanding, interior design acts as a densely structured reasoning problem that requires a model to simultaneously respect strict physical boundaries and nuanced aesthetic principles \cite{chen2024spatialvlm,hong20233d,farshad2023scenegenie}. To uncover the theoretical origins of their reasoning bottlenecks in such specialized micro-scenarios, we design targeted zero-shot evaluations and white-box probing experiments.

We first construct a evaluation subset from Structured3D \cite{zheng2020structured3d}, 3D-FRONT \cite{fu20213d}, and DesignHelper \cite{geng2025diffdesign} (\textbf{Figure \ref{fig:motivation}(a)}). This subset focuses on topological boundary cases (e.g., extremely narrow furniture spacing) and spaces with extreme color contrasts. We evaluate the performance of mainstream MLLMs, including GPT-4V and LLaVA-v1.6-7B, using standard layout evaluation metrics, including spatial relation accuracy (SRA) and collision detection error rate (CDER). The quantitative results in \textbf{Figure \ref{fig:motivation}(b)} reveal a severe performance degradation in boundary scenarios. For instance, while baseline models easily recognize loose spatial relations (e.g., predicting ``a table is in the center of the room'' with over 85\% accuracy), their CDER surges, often exceeding 40\%, when evaluating micro-topologies, such as determining whether a chair leg structurally intersects with a rug boundary within a narrow 5-pixel margin.

To architecturally deconstruct this failure mechanism, we employ Qwen2-VL-7B as a white-box probe. We extract the visual token attention heatmaps during the autoregressive decoding of specific spatial constraint instructions, such as ``check if the sofa collides with the back wall''.  Coupled with t-SNE dimensionality reduction, the visualizations expose a pronounced modality misalignment (\textbf{Figure \ref{fig:motivation}(c)}). For example, we observe that the model's visual attention weights remain highly concentrated on the global geometric centroids of the sofa and the wall to capture broad semantics. Conversely, the attention activations at the actual intersection contour, i.e., the critical boundary pixels that mathematically dictate physical collision, are extremely sparse or drop to near zero.

These empirical observations unequivocally demonstrate that the main cause of MLLM failure in interior design tasks lies in inadequate visual perceptual granularity. Because mainstream visual encoders (e.g., CLIP-ViT) are primarily optimized for global image-text semantic alignment during pre-training, they inevitably discard high-frequency local physical details (such as precise bounding box distance vectors) and fine-grained aesthetic color gamut distributions. When these low-resolution visual features are projected into the language model's latent space, the autoregressive decoder loses the precise visual grounding required for strict rule validation. Consequently, this leads to severe spatial hallucinations and aesthetic misjudgments. These results also explicitly underscore the necessity of introducing deterministic weak experts and correcting the latent trajectory, inspiring our methodology.

\section{Methodology}
\label{sec:method}

Based on the above analyses, we propose the Dual-prior Activation Residual Task-vectors Injection (DART-I) for MLLMs. Concurrent methods rely on scaling up training data for fine-tuning \cite{chen2025mindgpt,mao2025spatiallm,yang2025optiscene}, which incurs prohibitive computational costs and catastrophic forgetting. DART-I bypasses this expensive route via direct latent intervention. The core idea is to leverage training-free weak experts to extract spatial and aesthetic priors, and then use lightweight trainable modules to inject them into a frozen MLLM. Specifically, this framework operates through three steps: (i) extracting continuous priors (\textbf{Subsection \ref{sec:method_1_prior_extraction}}); (ii) projecting these priors into task vectors (\textbf{Subsection \ref{sec:method_2_task_vector}}); and (iii) dynamically injecting them as residual terms into the hidden states of MLLMs (\textbf{Subsection \ref{sec:method_3_injection}}), thus guiding the model towards precise reasoning for interior design. In \textbf{Subsection \ref{sec:methods_4_training}}, we briefly introduce how to optimize the lightweight trainable modules.
The framework is shown in \textbf{Figure \ref{fig:framework}}, with pseudo-code in \textbf{Appendix B}.

\subsection{Weak-to-Strong Prior Extraction}
\label{sec:method_1_prior_extraction}
In this step, we aim to recover the high-frequency local topological details that are typically discarded by MLLMs (demonstrated in \textbf{Subsection \ref{sec:empirical_evidence}}). To achieve this without introducing heavy costs, we introduce two non-parametric perception operators as our ``weak experts'', which respectively extract spatial and aesthetic priors to help ground the precise reasoning of MLLMs for interior design.

For the spatial dimension, we utilize phase congruency (PC), a frequency-domain measure that is fundamentally invariant to image contrast and brightness, to extract robust micro-topologies \cite{kovesi2000phase,fan2022phase}. The phase congruency map $PC(x,y)$ is calculated across multiple orientations and scales using Log-Gabor filters. Based on this robust boundary map, we compute the Geodesic Distance Transform (GDT) \cite{carlinet2026leveraging}, which measures the shortest path restricted to the image's spatial manifold, precisely encoding the pixel-level collision boundaries and layout spacing. Let $S(x,y)$ denote the binary contour map derived from thresholding $PC(x,y)$. The continuous distance map $D_{geo} \in \mathbb{R}^{H \times W}$ is formulated as:
\begin{equation}
D_{geo}(p) = \min_{q \in S} \inf_{\Gamma \in \mathcal{P}(p,q)} \int_{0}^{1} \left| \nabla I_v(\Gamma(t)) \right| dt,
\end{equation}
where $\mathcal{P}(p,q)$ denotes the set of all paths between pixel $p$ and boundary $q$. $\Gamma$ represents a parameterized path defined by the variable $t \in [0, 1]$, and $\nabla I_v$ is the spatial gradient of the original image. 
To convert this high-resolution manifold map into a compact continuous feature vector, we apply a spatial grid average-pooling operation over $M \times M$ non-overlapping patches, yielding the spatial prior vector $P_{spa} = \text{Pool}_{M \times M}(D_{geo}) \in \mathbb{R}^{M^2}$.

Simultaneously, to capture the nuanced visual aesthetic principles, we extract the fine-grained perceptual harmony and structural layout complexities. Since aesthetic harmony is highly sensitive to perceptual color shifts, we map $I_v$ into the perceptually uniform CIELAB color space. We compute the continuous color distribution moments (mean, variance, and skewness) across the $L^*, a^*, b^*$ channels to form a color representation vector $F_{color} \in \mathbb{R}^9$. Furthermore, to mathematically quantify the typographical balance and visual weight which MLLMs frequently misjudge, we perform a 2-level Discrete Wavelet Transform (DWT) on the luminance channel. We concatenate the normalized spectral energy of the high-frequency sub-bands (Horizontal, Vertical, and Diagonal) to represent the structural aesthetic complexity $F_{freq} \in \mathbb{R}^{3}$:
\begin{equation}
F_{freq}^{(i)} = \frac{1}{H_i W_i} \sum_{x,y} \left| W_{\psi}^{(i)}(x,y) \right|^2, \quad i \in \{H, V, D\},
\end{equation}
where $W_{\psi}^{(i)}$ denotes the wavelet coefficients of the corresponding high-frequency sub-band, and $H_i, W_i$ are the height and width of that sub-band. The aesthetic prior vector is thus a fusion of perceptual color and structural frequency:
\begin{equation}
P_{aes} = [F_{color} \parallel F_{freq}] \in \mathbb{R}^{12}.
\end{equation}
Finally, these two representations are concatenated to form the dual-prior constraint vector $P \in \mathbb{R}^{M^2 + 12}$, formulated as $P = [P_{spa} \parallel P_{aes}]$.

\subsection{Prior-aware Semantic Translation}
\label{sec:method_2_task_vector}

Although the extracted operator priors are informative, they reside in analytical feature spaces and cannot be directly injected into the latent semantic space of the MLLM. To bridge this gap, we introduce a lightweight semantic translation module that maps each operator output into the hidden dimension of the backbone model.

Specifically, let $\mathbf{X}_{\mathrm{vis}} \in \mathbb{R}^{N \times d_{\mathrm{v}}}$ denote the visual tokens from the frozen vision encoder. We first extract a global scene descriptor $\mathbf{g} = \text{AvgPool}(\mathbf{X}_{\mathrm{vis}})$. To enable the model to adaptively scale the contribution of structural vs. aesthetic cues (e.g., a cluttered room vs. a minimalist color palette), we apply a vision-conditioned weighting mechanism, which can be expressed as:
\begin{equation}
\tilde{P} = \sigma \left( \mathbf{W}_{\mathrm{g}} \mathbf{g} + \mathbf{b}_{\mathrm{g}} \right) \odot [P_{\mathrm{spa}} \parallel P_{\mathrm{aes}}],
\end{equation}
where $\sigma(\cdot)$ is the Sigmoid function, $\odot$ denotes the element-wise product, and $\mathbf{W}_{\mathrm{g}}, \mathbf{b}_{\mathrm{g}}$ are trainable parameters. This step ensures that the weak priors do not indiscriminately override the model's inherent perceptions but rather augment them where the visual context suggests high uncertainty.
The fused prior $\tilde{P}$ is then translated into a directional task vector $V_{\mathrm{task}}$ through:
\begin{equation}
V_{\mathrm{task}} = \text{GeLU}(W_{\mathrm{proj}} \tilde{P} + b_{\mathrm{proj}}),
\end{equation}
where $W_{\mathrm{proj}}$ maps the scene-conditioned prior into the hidden dimension $d$ of the MLLM. $V_{\mathrm{task}}$ now represents a steering intent in the latent space, ready to guide the reasoning trajectory.

\subsection{Dynamic Residual Latent Injection}
\label{sec:method_3_injection}
The final stage of the DART-I is the dynamic injection of the projected task vector $V_{task}$ into the internal forward pass of the MLLM. As established in our problem formulation, standard zero-shot inference relies on unguided hidden states, which inevitably drift towards hallucinated spatial topologies when confronted with dense design constraints. Our objective is to steer the reasoning trajectory at the fundamental logical level by explicitly anchoring these hidden states with our physical and aesthetic task vectors.
Besides, given that not all layers or tokens in an autoregressive sequence require design-rule enforcement, an indiscriminate injection would impair the model's linguistic fluency. We therefore propose a dynamic residual mechanism that modulates the intervention at both the layer and token levels. This ensures that the design priors are only active when and where they are relevant to the reasoning task.

\paragraph{Layer-wise Routing.} Since different transformer layers handle varying degrees of perception and logic, we estimate a routing score $r^{(l)}$ for each target layer $l$. This score determines the necessity of intervention at a specific depth of the network:
\begin{equation}
r^{(l)} = \sigma \left( W_{\mathrm{layer}}^{(l)} [\bar{h}^{(l)} \parallel V_{\mathrm{task}}] + b_{\mathrm{layer}}^{(l)} \right),
\end{equation}
where $\bar{h}^{(l)}$ is the average-pooled representation of the current layer.

\paragraph{Token-wise Gating.} Furthermore, during the autoregressive generation of an evaluation report, the importance of design constraints fluctuates. For instance, generating the word ``is'' requires less guidance than generating ``collision''. We thus predict an instance-specific gating scalar $\alpha_t^{(l)}$ for each token $t$:
\begin{equation}
\alpha_t^{(l)} = \sigma \left( W_{\mathrm{gate}}^{(l)} [h_t^{(l)} \parallel V_{\mathrm{task}}] + b_{\mathrm{gate}}^{(l)} \right).
\end{equation}
The final intervened hidden state $\hat{h}_t^{(l)}$ is computed through an adaptive residual connection:
\begin{equation}
\hat{h}_t^{(l)} = h_t^{(l)} + r^{(l)} \cdot \alpha_t^{(l)} \cdot V_{\mathrm{task}}.
\end{equation}
By dynamically nudging the latent trajectory, DART-I ensures that the MLLM's final probability distribution $p_\theta(y_t \mid I_v, y_{<t})$ is anchored by deterministic physical and aesthetic boundaries. This architecture achieves expert-level reasoning in the interior design domain while completely bypassing the alignment tax of fine-tuning and preserving the model's general intelligence.

\subsection{Overall Objective}
\label{sec:methods_4_training}
To optimize the trainable modules, we use a standard causal language modeling task. Given the input image $I_v$ and the ground-truth text sequence $y = \{y_1, y_2, \dots, y_T\}$, the model minimizes the auto-regressive cross-entropy loss, expressed as:
\begin{equation}
    \mathcal{L}(\Theta_{\mathrm{DART}}) = - \sum_{t=1}^{T} \log p_\theta \left(y_t \mid I_v, y_{<t}; \Theta_{\mathrm{DART}} \right)
\end{equation}
where $\Theta_{\mathrm{DART}}$ denotes the trainable parameters of the translation and injection modules (i.e., $\mathbf{W}_{\mathrm{g}}, W_{\mathrm{proj}}, \mathbf{W}_{\mathrm{layer}}^{(l)}, \mathbf{W}_{\mathrm{gate}}^{(l)}$). The backbone MLLM remains strictly frozen. Gradients backpropagate only to update $\Theta_{\mathrm{DART}}$. By minimizing this generation loss, the routing and gating modules learn to adjust the intervention dynamically, mapping the extracted priors to the appropriate text tokens.

\section{Theoretical Analysis}
\label{sec:theoretical}
To justify the efficacy of DART-I, we provide a theoretical analysis demonstrating how our dynamic residual latent injection bounds the probability of generating hallucinations. In MLLMs, the final next-token probability distribution is derived by projecting the last layer's hidden state $h^{(L)}$ into the vocabulary space $\mathcal{V}$ via the unembedding matrix $W_U \in \mathbb{R}^{|\mathcal{V}| \times d}$. We partition $\mathcal{V}$ into a feasible token set $\mathcal{V}_{fea}$ (i.e., tokens adhering to physical and aesthetic constraints) and a hallucination token set $\mathcal{V}_{hal}$ (i.e., tokens violating constraints). The core premise of DART-I is that the injected directional task vector $V_{task}$ acts as an anchor, shifting the latent representations towards the feasible constraint manifold. Thus, we get:
\begin{theorem}\label{theorem:1}
    Let $W_y \in \mathbb{R}^d$ be the row vector in $W_U$ corresponding to token $y$. Assume the projection network maps the priors such that $V_{task}$ achieves an alignment margin $\Delta > 0$ separating feasible design tokens from hallucinated tokens, i.e., $\min_{y \in \mathcal{V}_{fea}} W_y V_{task} - \max_{y' \in \mathcal{V}_{hal}} W_{y'} V_{task} \geq \Delta$. Under the assumption of approximate linear propagation in residual streams, the probability of generating a spatial hallucination is exponentially bounded by:
\begin{equation}
    P_{DART}(\hat{y} \in \mathcal{V}_{hal}) \leq P_{base}(\hat{y} \in \mathcal{V}_{hal}) \exp(-\alpha \Delta), 
\end{equation}
    where $P*{base}$ is the unguided generation probability of the frozen MLLM, and $\alpha$ is the effective intervention strength.
\end{theorem}
\textbf{Theorem \ref{theorem:1}} shows that DART-I suppresses spatial hallucinations by establishing an exponential upper bound. It demonstrates that as long as the injected task vector explicitly separates feasible tokens from hallucinated ones by a positive margin $\Delta$, the probability of generating invalid spatial topologies decreases exponentially. The original hallucination rate of the MLLM is bounded and controlled by both the intervention strength $\alpha$ and the feature margin $\Delta$.
The detailed proofs and analyses are provided in \textbf{Appendix A}.

\begin{table*}[t]
\centering
\caption{Quantitative performance on ArchitecBench and MMBench. We report the mean metrics along with the 95\% confidence intervals (subscript) across 5 random seeds. The best results are highlighted in bold. More results are provided in Appendix F.}\vspace{-0.1in}
\label{tab:main_results}
\resizebox{0.9\textwidth}{!}{
\begin{tabular}{lcccccc}
\toprule
\textbf{Model} & \textbf{Paradigm} & \textbf{CDER (\%)} $\downarrow$ & \textbf{SRA (\%)} $\uparrow$ & \textbf{FPS (\%)} $\uparrow$ & \textbf{AAS (\%)} $\uparrow$ & \textbf{MMBench} $\uparrow$ \\
\midrule
InstructBLIP-Vicuna-7B & Zero-shot & $66.1_{\pm 1.0}$ & $40.8_{\pm 2.1}$ & $38.2_{\pm 2.3}$ & $37.9_{\pm 1.2}$ & $60.5$ \\
LLaVA-1.5-7B & Zero-shot & $61.8_{\pm 2.4}$ & $46.5_{\pm 1.9}$ & $44.7_{\pm 2.1}$ & $42.5_{\pm 1.8}$ & $64.3$ \\
LLaVA-1.5-13B & Zero-shot & $56.4_{\pm 1.9}$ & $51.2_{\pm 2.5}$ & $49.1_{\pm 1.8}$ & $47.3_{\pm 1.6}$ & $67.7$ \\
LLaVA-NeXT-34B & Zero-shot & $48.2_{\pm 1.5}$ & $60.3_{\pm 1.1}$ & $57.5_{\pm 1.3}$ & $54.9_{\pm 1.2}$ & $79.3$ \\
Qwen2-VL-7B & Zero-shot & $52.5_{\pm 1.8}$ & $52.7_{\pm 1.6}$ & $50.1_{\pm 1.2}$ & $50.4_{\pm 1.5}$ & $81.0$ \\
Qwen2-VL-72B & Zero-shot & $38.5_{\pm 1.1}$ & $66.4_{\pm 0.9}$ & $65.1_{\pm 1.4}$ & $70.2_{\pm 0.9}$ & 85.7 \\
\midrule
LLaVA-1.5-7B + LoRA & Parameter-Efficient Fine-Tuning & $28.4_{\pm 0.8}$ & $74.5_{\pm 0.7}$ & $73.6_{\pm 0.9}$ & $75.4_{\pm 0.8}$ & $58.2 \textit{ (Drop)}$ \\
Qwen2-VL-7B + LoRA & Parameter-Efficient Fine-Tuning & $27.1_{\pm 0.7}$ & $75.5_{\pm 0.6}$ & $75.8_{\pm 0.8}$ & $77.1_{\pm 0.7}$ & $72.5 \textit{ (Drop)}$ \\
\midrule
LLaVA-1.5-7B + DART-I & Latent Injection & $26.1_{\pm 0.5}$ & $77.2_{\pm 0.4}$ & $74.8_{\pm 0.5}$ & $79.5_{\pm 0.3}$ & $64.1 \textit{ (Retained)}$\\
LLaVA-1.5-13B + DART-I & Latent Injection & $25.2_{\pm 1.1}$ & $78.6_{\pm 0.5}$ & $77.1_{\pm 0.4}$ & $81.1_{\pm 0.7}$ & $67.6 \textit{ (Retained)}$\\
Qwen2-VL-7B + DART-I & Latent Injection & $\mathbf{24.1_{\pm 0.8}}$ & $\mathbf{79.4_{\pm 0.4}}$ & $\mathbf{80.3_{\pm 0.9}}$ & $\mathbf{81.7_{\pm 0.3}}$ & $80.8 \textit{ (Retained)}$\\
\bottomrule
\end{tabular}
}
\end{table*}

\begin{figure}
    \centering
    \includegraphics[width=\columnwidth]{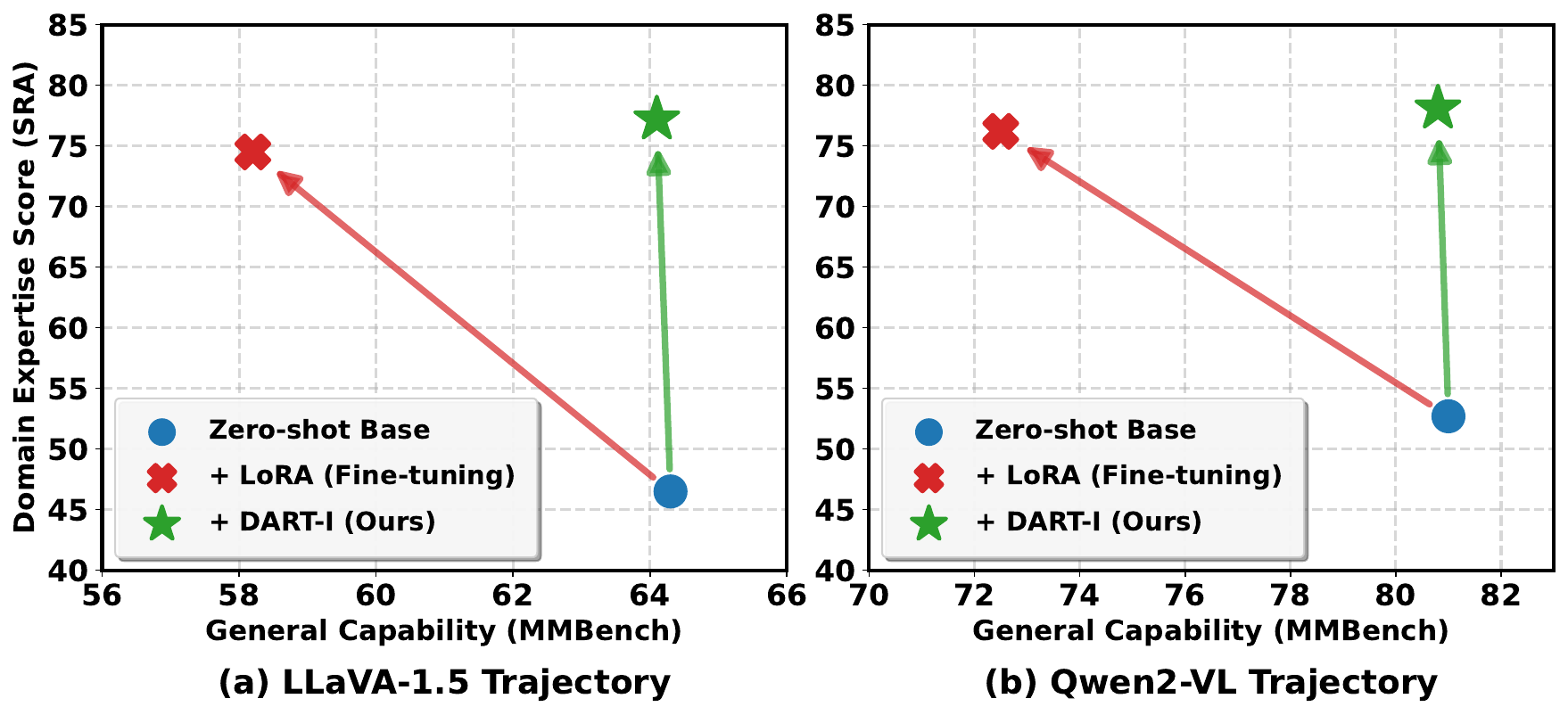}
    \caption{Trade-off analysis on different base models.}
    \label{fig:tradeoff}
\end{figure}

\begin{figure}
    \centering
    \includegraphics[width=\columnwidth]{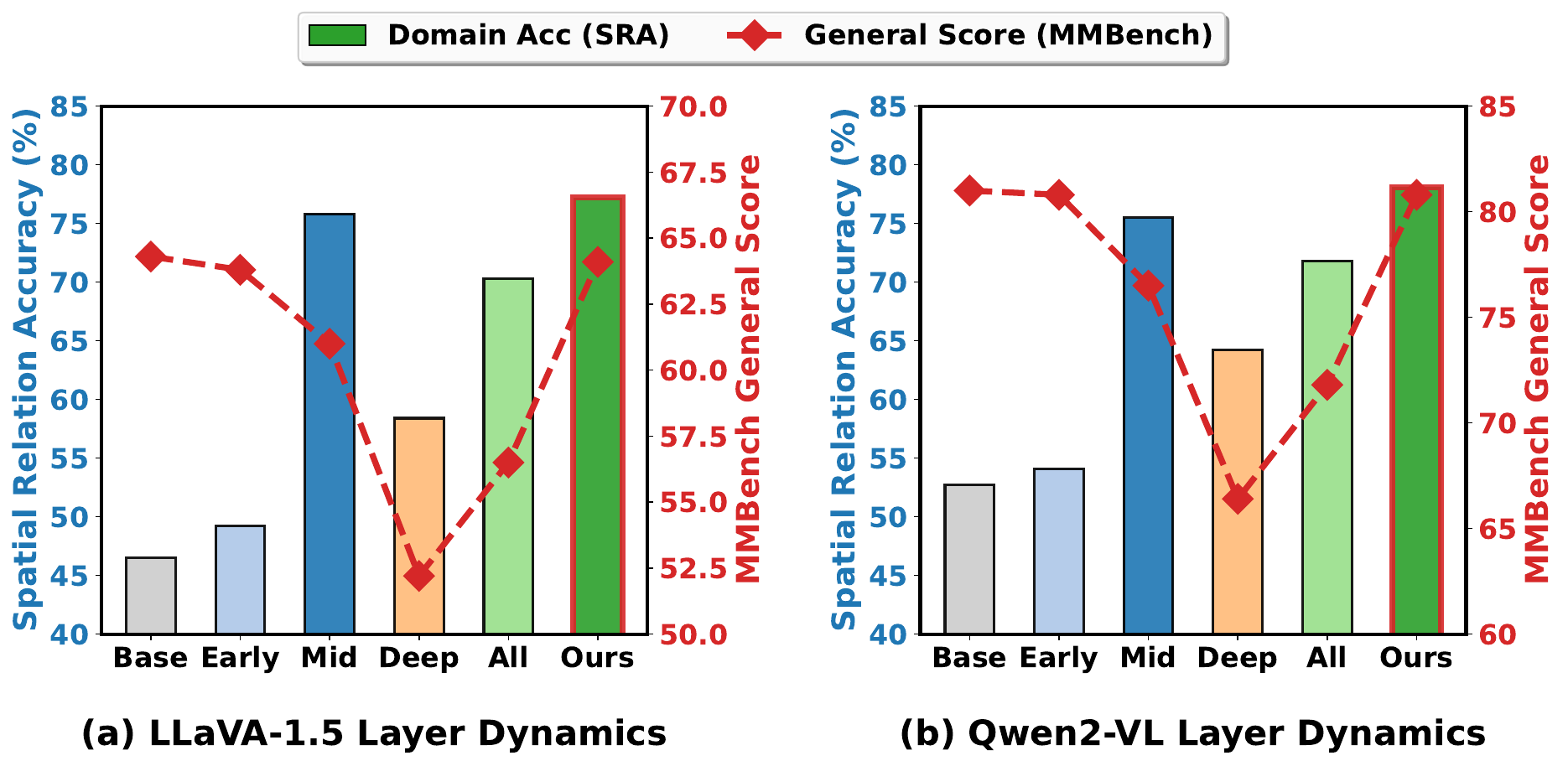}
    \caption{Performance of different base models across different intervention layer subsets. See Appendix G for details.}
    \label{fig:layer_dynamics}
    \vspace{-0.2in}
\end{figure}

\section{Experiments}
\label{sec:experiment}
In this section, we conduct extensive experiments on various benchmark datasets to evaluate the effectiveness of our method. More details and experiments are provided in \textbf{Appendices C-F}.

\subsection{Experimental Setup}
\label{sec:exp_settings}
To evaluate MLLMs in highly constrained interior design scenarios, we develop ArchitecBench, a specialized benchmark integrating complex indoor and outdoor architecture environments with expert annotations from Structured3D \cite{zheng2020structured3d}, Open3D-VQA \cite{zhang2025open3d}, IL3D \cite{zhou2025il3d}, 3D-FRONT \cite{fu20213d}, and DesignHelper \cite{geng2025diffdesign}. It comprises 2,500 densely constrained instances, categorized into a spatial topology subset for assessing micro-topologies, such as physical collisions and furniture spacing, and an aesthetic harmony subset for evaluating sensitivity to color contrast, visual weight, and layout balance. We evaluate DART-I with twelve representative MLLMs, including closed-source frontiers like GPT-4o, Gemini 1.5, and Claude 3.5 Sonnet, alongside open-source baselines across various scales, e.g., InstructBLIP, LLaVA-1.5/NeXT, and Qwen2-VL. To demonstrate the advantages of our training-free intervention over traditional paradigms, we further establish domain-specific fine-tuning baselines by applying low-rank adaptation (LoRA) \cite{hu2022lora} to LLaVA-1.5-7B and Qwen2-VL-7B using 50,000 interior design instruction pairs. The performance is quantified via five primary metrics: collision detection error rate (CDER), spatial relation accuracy (SRA), and functional planning score (FPS) for spatial reasoning; the aesthetic alignment score (AAS), measuring Pearson correlation with expert ratings for aesthetic evaluation; and MMBench \cite{liu2024mmbench} to monitor potential catastrophic forgetting of general capabilities. All experimental results are reported with 95\% confidence intervals derived from five independent runs with different random seeds. More details and implementations are provided in \textbf{Appendices C-F}.

\subsection{Quantitative Performance}
\label{sec:performance_comparison}
The quantitative results across all baselines and our DART-I variants are summarized in \textbf{Table \ref{tab:main_results}}. The empirical observations indicate that existing MLLMs struggle when confronted with interior design tasks. While scaling up the parameters in open-source models provides consistent but marginal improvements, they fall short of professional engineering standards and exhibit high generation variance ($\pm 1.5$ to $\pm 2.4$). Standard LoRA fine-tuning substantially improves domain-specific metrics by reducing the CDER of LLaVA-1.5-7B and increasing its SRA. 
Conversely, the integration of DART-I into the frozen backbones achieves highly competitive reasoning performance without any domain-specific training. By directly injecting deterministic analytical priors into the latent space, DART-I reduces the CDER of Qwen2-VL-7B and elevates the FPS. This performance significantly outperforms the fine-tuned LoRA baselines and is highly comparable to leading closed-source models, closely approaching the ceiling set by Gemini 1.5 Pro. Notably, the confidence intervals for DART-I are relatively narrow compared to standard autoregressive generation. This explicitly proves that anchoring latent states with deterministic mathematical priors fundamentally stabilizes the reasoning variance and prevents hallucinatory drift.

\subsection{Trade-off Analysis}
\label{sec:tradeoff_analysis}
A critical challenge in adapting foundational models to vertical domains like architectural design is the occurrence of catastrophic forgetting. As observed in the MMBench scores (\textbf{Table \ref{tab:main_results}}), standard LoRA fine-tuning irreparably damages the model's general common sense, causing LLaVA-1.5-7B's general capability score to plummet from $64.3$ to $58.2$. 
 To visualize this limitation, we plot the Pareto frontier in \textbf{Figure \ref{fig:tradeoff}}. The horizontal axis represents the general capability (MMBench), and the vertical axis represents domain expertise (SRA). The fine-tuned models exhibit a severe trade-off, moving towards the top-left quadrant of the plot. Conversely, because DART-I strictly freezes the MLLM backbone and only injects residual perturbations dynamically, it completely bypasses catastrophic forgetting. All models equipped with DART-I move straight vertically along the domain-accuracy axis, demonstrating a strict Pareto improvement and establishing our method as a non-destructive domain adaptation paradigm.

\subsection{Human Judgment and Expert Alignment}
\label{sec:human_judgment}
Since automated metrics often fail to fully capture the nuanced expertise required in real-world design, we conduct a human judgment collection to evaluate model-to-human alignment. We recruit 20 professional interior designers (each possessing over 5 years of industry experience) to participate in a blind A/B/C preference testing protocol. The experts evaluate the output rationales generated by GPT-4o, Gemini 1.5 Pro, LLaVA-1.5-7B + LoRA, and LLaVA-1.5-7B + DART-I across 500 challenging scenes. The designers rank the outputs based on structural legality, spatial logical soundness, and aesthetic professionality, entirely blind to the model identities.
The human preference outcomes are presented in \textbf{Table \ref{tab:human_eval}}, utilizing the Elo rating system and pairwise win rates against the LoRA baseline. Gemini 1.5 Pro establishes the highest Elo rating of $1120$. DART-I follows closely behind with an Elo rating of $1095$, successfully surpassing the fine-tuned LoRA baseline ($1012$) and outperforming GPT-4o ($1068$). In direct pairwise comparisons against the LoRA baseline, DART-I secures a reliable $62.5\%$ win rate. 
More qualitative comparison and analyses are provided in \textbf{Appendix F}.

\subsection{Intervention Layer Dynamics}
\label{sec:layer_dynamics}
To understand how latent intervention impacts different reasoning stages, we evaluate DART-I by injecting the task vectors into varying subsets of transformer layers $\mathcal{L}$. The 32 layers of LLaVA-1.5-7B are grouped into Early (0-10), Middle (11-21), and Deep (22-31) blocks. The results are shown in \textbf{Figure \ref{fig:layer_dynamics}}. Injecting priors exclusively into the early layers yields marginal domain improvements, as these shallow blocks primarily process low-level visual features and basic linguistic syntax. Injecting exclusively into the deepest layers degrades grammatical coherence and negatively impacts MMBench scores, because deep layers are highly specialized for final vocabulary projection. The optimal performance emerges exclusively when interventions target the Middle-to-Deep transition blocks (layers 15-25). This empirical finding confirms that middle layers encode the core semantic routing and causal logic within MLLMs, making them the optimal semantic space for grounding external physical and aesthetic constraints.

\begin{table}
\centering
\caption{Human expert evaluation results on 500 densely constrained scenes.}\vspace{-0.1in}
\label{tab:human_eval}
\resizebox{\columnwidth}{!}{
\begin{tabular}{lccc}
\toprule
\textbf{Model} & \textbf{Elo Rating} $\uparrow$ & \textbf{Win (\%)} $\uparrow$ & \textbf{Tie (\%)} \\
\midrule
GPT-4o & $1068$ & $54.2$ & $12.5$ \\
Gemini 1.5 Pro & $1120$ & $68.4$ & $15.2$ \\
LLaVA-1.5-7B (Zero-shot) & $815$ & $11.8$ & $4.6$ \\
LLaVA-1.5-7B + LoRA & $1012$ & - & - \\
LLaVA-1.5-7B + DART-I & 1095 & 62.5 & 16.4 \\
\bottomrule
\end{tabular}
}
\end{table}

\begin{table}[t]
\centering
\caption{Model reasoning accuracy (\%) under varying numbers of simultaneous design constraints.}\vspace{-0.1in}
\label{tab:constraint_density}
\resizebox{\columnwidth}{!}{
\begin{tabular}{lcccc}
\toprule
\textbf{Model} & \textbf{1 Const.} & \textbf{2 Const.} & \textbf{3 Const.} & \textbf{5 Const.} \\
\midrule
Gemini 1.5 Pro & $83.5$ & $76.1$ & $69.2$ & $57.8$ \\
LLaVA-1.5-7B & $57.6$ & $39.8$ & $29.1$ & $21.3$ \\
LLaVA-1.5-7B + LoRA & $80.2$ & $71.7$ & $55.4$ & $41.6$ \\
LLaVA-1.5-7B + DART-I & 81.4 & 75.3 & 66.8 & 53.9 \\
\bottomrule
\end{tabular}
}
\end{table}

\begin{table}
\centering
\caption{Computational overhead profiling for adaptation.}\vspace{-0.1in}
\label{tab:overhead}
\resizebox{\columnwidth}{!}{
\begin{tabular}{lccc}
\toprule
\textbf{Method} & \textbf{Extra Params} & \textbf{VRAM Usage} & \textbf{Latency} \\
\midrule
LoRA Fine-tuning & 19.8 M & 18.5 GB & + 1.2 ms/tok \\
\textbf{Ours} & \textbf{0.6 M} & \textbf{4.2 GB} & \textbf{+ 0.11 ms/tok} \\
\bottomrule
\end{tabular}
}
\end{table}

\begin{figure}[t]
    \centering
    \includegraphics[width=\columnwidth]{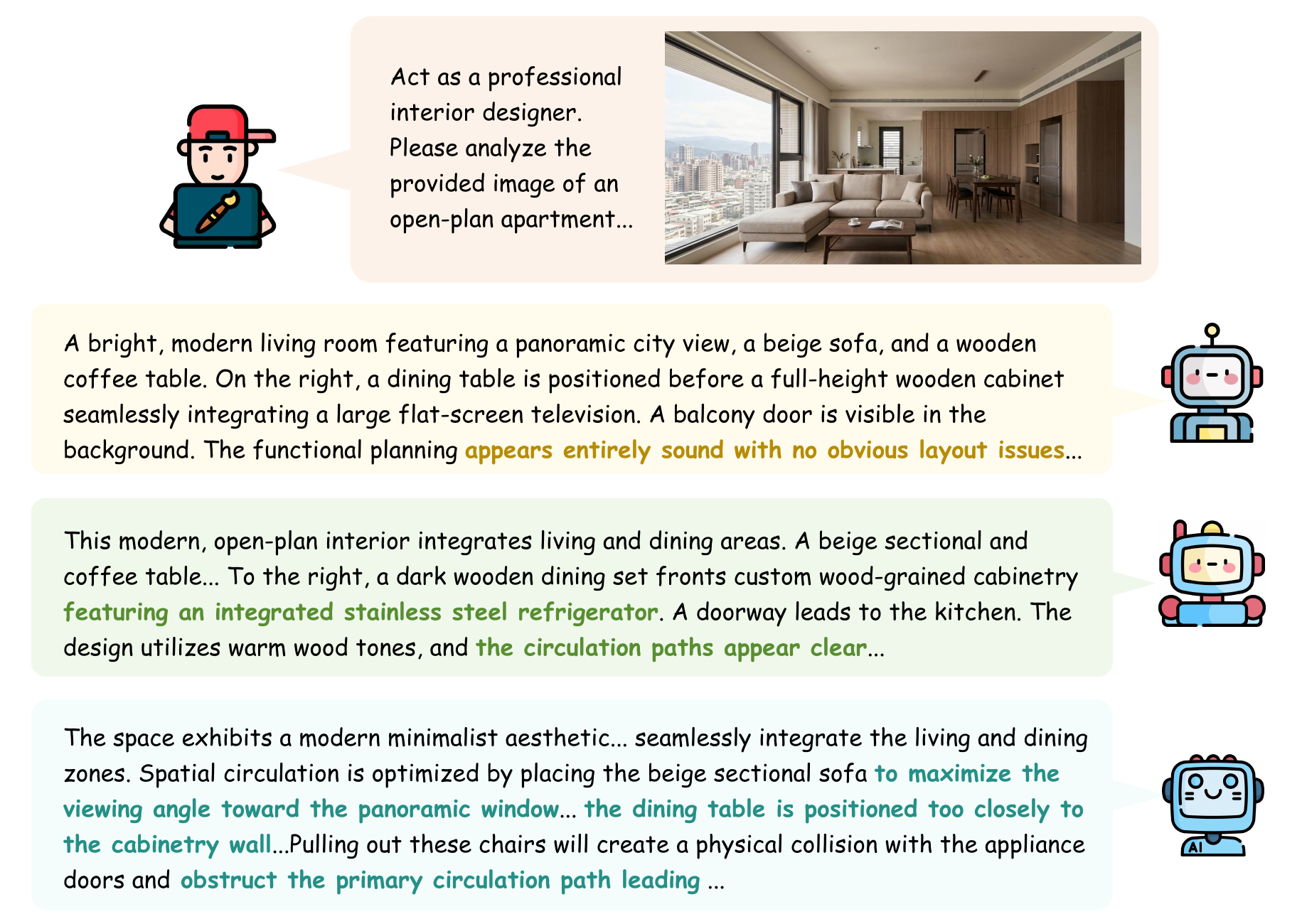}\vspace{-0.1in}
    \caption{Visualization of different methods.}
    \label{fig:vis}
\end{figure}

\begin{table}[t]
\centering
\caption{Ablation studies of the DART-I components.}
\vspace{-0.1in}
\label{tab:ablation}
\resizebox{\columnwidth}{!}{
\begin{tabular}{lcccc}
\toprule
\textbf{Configuration} & \textbf{CDER} $\downarrow$ & \textbf{SRA} $\uparrow$ & \textbf{AAS} $\uparrow$ & \textbf{MMBench} $\uparrow$ \\
\midrule
Base Model & 61.8 & 46.5 & 42.5 & 64.3 \\
+ Spatial Prior ($P_{spa}$) & 35.7 & 71.8 & 51.4 & 63.9 \\
+ Aesthetic Prior ($P_{aes}$) & 58.2 & 49.1 & 73.5 & 64.1 \\
+ Both Priors (Fixed $\alpha=0.5$) & 28.6 & 74.9 & 76.2 & 62.8 \\
+ \textbf{Full DART-I (Adaptive $\alpha_t^{(l)}$)} & 26.1 & 77.2 & 79.5 & 64.1 \\
\bottomrule
\end{tabular}
}
\end{table}

\subsection{Robustness against Constraint Density}
\label{sec:robustness}
Interior design instructions often contain dense, multi-hop constraints. We evaluate the robustness of the models by incrementally increasing the number of simultaneous constraints per prompt from 1 to 5. As presented in \textbf{Table \ref{tab:constraint_density}}, the base LLaVA-1.5-7B experiences a catastrophic cognitive collapse under increasing load, with its overall accuracy plummeting from $57.6\%$ to $21.3\%$ when handling 5 simultaneous constraints. The fine-tuned LoRA baseline mitigates this issue initially but still suffers a severe accuracy drop to $41.6\%$. Conversely, DART-I exhibits superior robustness, e.g., it maintains an accuracy of $53.9\%$ even under 5 simultaneous constraints. This demonstrates its effectiveness in multi-conditional scenarios.

\subsection{Computational Efficiency Profiling}
\label{sec:computational_overhead}
We evaluate the computational overhead of our method compared to other baselines in \textbf{Table \ref{tab:overhead}}. For a single card of our cluster under 4-bit quantization, standard fine-tuning through LoRA requires updating approximately 19.8 million parameters and consuming 18.5 GB of training VRAM; while DART-I updates only the linear projection network and the gating module, representing less than $0.01\%$ of the backbone and requires only 4.2 GB of VRAM of a single card. During inference, the weak experts operate purely in the pre-filling stage, introducing a latency overhead of merely 0.11 ms/token, further demonstrating the advantages of DART-I.

\subsection{Visualization}
To qualitatively assess the effectiveness of DART-I in understanding and analyzing complex interior design scenes, we present a series of visualizations comparing three variants based on Qwen2-VL-7B: (i) a general-purpose base model, (ii) a LoRA fine-tuned model, and (iii) model with DART-I. An example of visualization results is shown in \textbf{Figure \ref{fig:vis}}. It provides a design with a modern, open-plan apartment featuring distinct living and dining areas, characterized by a panoramic window with a city skyline, a beige sectional sofa, and a seamless wall of custom wood-grained cabinetry with integrated stainless steel appliances. We can observe that the base model primarily produces descriptions with spatial and object-identity hallucinations, such as misidentifying a built-in appliance. The LoRA-tuned model improves accuracy but remains largely descriptive, focusing on item identification without professional depth. In contrast, our method accurately reasons about spatial relationships, analyzes the function of bespoke cabinetry and panoramic windows, and evaluates the overall modern minimalist aesthetic. The evaluation with human experts in \textbf{Subsection \ref{sec:human_judgment}} also verifies this. More results are provided in \textbf{Appendix F}.

\subsection{Ablation Studies}
\label{sec:ablation_studies}
To deconstruct the specific contribution of each module, we conduct component ablation studies on the LLaVA-1.5-7B backbone, as detailed in \textbf{Table \ref{tab:ablation}}. The baseline model lacking any priors suffers from severe modality misalignment. Integrating only the spatial prior effectively drops the CDER to $35.7\%$, proving that extracting micro-topologies prevents physical spatial collisions. However, without the aesthetic prior, the AAS remains a suboptimal $51.4\%$. Integrating both priors with a fixed injection strength ($\alpha = 0.5$) significantly improves domain metrics but introduces a slight degradation in linguistic fluency (MMBench drops to $62.8$). Ultimately, utilizing the full DART-I mechanism with the data-dependent adaptive gating module achieves the optimal configuration. The gating mechanism routes the intervention strength token-by-token, perfectly preserving the language model's inherent reasoning fluency while maximizing physical constraint alignment.

\section{Conclusion}
\label{sec:conclusion}
In this paper, we identify a critical modality misalignment issue in MLLMs when applied to densely constrained interior design tasks. Our empirical analyses reveal that the loss of high-frequency local topological details and aesthetic shifts during visual encoding deprives MLLMs of precise visual grounding, leading to hallucinations. To overcome this generalization bottleneck, we propose DART-I, a Dual-prior Activation Residual Task-vectors Injection mechanism for MLLMs. Operating under a novel weak-to-strong paradigm, DART-I explicitly extracts continuous spatial and aesthetic priors via lightweight experts. By transforming these priors into directional task vectors and dynamically injecting them as residual terms into the latent space of frozen MLLMs, our framework guides the models towards precise reasoning for interior design. 
Extensive experiments demonstrate the advantages of DART-I, i.e., it effectively aligns evaluations with human experts while achieving superior zero-shot reasoning performance at a minimal computational cost.


\bibliographystyle{ACM-Reference-Format}
\bibliography{sample-base}

@book{ching2018interior,
  title={Interior design illustrated},
  author={Ching, Francis DK and Binggeli, Corky},
  year={2018},
  publisher={John Wiley \& Sons}
}

@inproceedings{ulrich1991effects,
  title={Effects of interior design on wellness: theory and recent scientific research.},
  author={Ulrich, Roger S},
  booktitle={Journal of Health Care Interior Design: Proceedings from the... Symposium on Health Care Interior Design. Symposium on Health Care Interior Design},
  volume={3},
  pages={97--109},
  year={1991}
}

@book{pile2005history,
  title={A history of interior design},
  author={Pile, John F},
  year={2005},
  publisher={Laurence King Publishing}
}

@inproceedings{nauata2020house,
  title={House-gan: Relational generative adversarial networks for graph-constrained house layout generation},
  author={Nauata, Nelson and Chang, Kai-Hung and Cheng, Chin-Yi and Mori, Greg and Furukawa, Yasutaka},
  booktitle={European Conference on Computer Vision},
  pages={162--177},
  year={2020},
  organization={Springer}
}

@article{hu2020graph2plan,
  title={Graph2plan: Learning floorplan generation from layout graphs},
  author={Hu, Ruizhen and Huang, Zeyu and Tang, Yuhan and Van Kaick, Oliver and Zhang, Hao and Huang, Hui},
  journal={ACM Transactions on Graphics (TOG)},
  volume={39},
  number={4},
  pages={118--1},
  year={2020},
  publisher={ACM New York, NY, USA}
}

@article{li2019layoutgan,
  title={Layoutgan: Generating graphic layouts with wireframe discriminators},
  author={Li, Jianan and Yang, Jimei and Hertzmann, Aaron and Zhang, Jianming and Xu, Tingfa},
  journal={arXiv preprint arXiv:1901.06767},
  year={2019}
}

@article{wang2019planit,
  title={Planit: Planning and instantiating indoor scenes with relation graph and spatial prior networks},
  author={Wang, Kai and Lin, Yu-An and Weissmann, Ben and Savva, Manolis and Chang, Angel X and Ritchie, Daniel},
  journal={ACM Transactions on Graphics (TOG)},
  volume={38},
  number={4},
  pages={1--15},
  year={2019},
  publisher={ACM New York, NY, USA}
}

@article{alayrac2022flamingo,
  title={Flamingo: a visual language model for few-shot learning},
  author={Alayrac, Jean-Baptiste and Donahue, Jeff and Luc, Pauline and Miech, Antoine and Barr, Iain and Hasson, Yana and Lenc, Karel and Mensch, Arthur and Millican, Katherine and Reynolds, Malcolm and others},
  journal={Advances in neural information processing systems},
  volume={35},
  pages={23716--23736},
  year={2022}
}

@inproceedings{li2023blip,
  title={Blip-2: Bootstrapping language-image pre-training with frozen image encoders and large language models},
  author={Li, Junnan and Li, Dongxu and Savarese, Silvio and Hoi, Steven},
  booktitle={International conference on machine learning},
  pages={19730--19742},
  year={2023},
  organization={PMLR}
}

@article{liu2023visual,
  title={Visual instruction tuning},
  author={Liu, Haotian and Li, Chunyuan and Wu, Qingyang and Lee, Yong Jae},
  journal={Advances in neural information processing systems},
  volume={36},
  pages={34892--34916},
  year={2023}
}

@article{ouyang2022training,
  title={Training language models to follow instructions with human feedback},
  author={Ouyang, Long and Wu, Jeffrey and Jiang, Xu and Almeida, Diogo and Wainwright, Carroll and Mishkin, Pamela and Zhang, Chong and Agarwal, Sandhini and Slama, Katarina and Ray, Alex and others},
  journal={Advances in neural information processing systems},
  volume={35},
  pages={27730--27744},
  year={2022}
}

@article{guo2025copo,
  title={COPO: Causal-Oriented Policy Optimization for Hallucinations of MLLMs},
  author={Guo, Peizheng and Wang, Jingyao and Qiang, Wenwen and Zhou, Jiahuan and Zheng, Changwen and Hua, Gang},
  journal={arXiv preprint arXiv:2508.04182},
  year={2025}
}

@article{geng2025diffdesign,
  title={DiffDesign: Controllable diffusion with meta prior for efficient interior design generation},
  author={Geng, Tao and Yang, Yuxuan},
  journal={PloS one},
  volume={20},
  number={9},
  pages={e0331240},
  year={2025},
  publisher={Public Library of Science San Francisco, CA USA}
}

@inproceedings{ccelen2024design,
  title={I-design: Personalized llm interior designer},
  author={{\c{C}}elen, Ata and Han, Guo and Schindler, Konrad and Van Gool, Luc and Armeni, Iro and Obukhov, Anton and Wang, Xi},
  booktitle={European Conference on Computer Vision},
  pages={217--234},
  year={2024},
  organization={Springer}
}

@article{wang2025learning,
  title={Learning to think: Information-theoretic reinforcement fine-tuning for llms},
  author={Wang, Jingyao and Qiang, Wenwen and Song, Zeen and Zheng, Changwen and Xiong, Hui},
  journal={arXiv preprint arXiv:2505.10425},
  year={2025}
}

@article{chen2025mindgpt,
  title={Mindgpt-4ov: An enhanced mllm via a multi-stage post-training paradigm},
  author={Chen, Wei and Du, Chaoqun and Gu, Feng and He, Wei and Li, Qizhen and Liu, Zide and Pan, Xuhao and Ren, Chang and Rao, Xudong and Wang, Chenfeng and others},
  journal={arXiv preprint arXiv:2512.02895},
  year={2025}
}

@phdthesis{gallega2025exploring,
  title={Exploring the Use of Generative AI for Material Selection in Interior Design},
  author={Gallega, Tan and Wharlo, Rgee},
  year={2025},
  school={Future University Hakodate}
}

@inproceedings{zheng2020structured3d,
  title={Structured3d: A large photo-realistic dataset for structured 3d modeling},
  author={Zheng, Jia and Zhang, Junfei and Li, Jing and Tang, Rui and Gao, Shenghua and Zhou, Zihan},
  booktitle={European Conference on Computer Vision},
  pages={519--535},
  year={2020},
  organization={Springer}
}

@inproceedings{fu20213d,
  title={3d-front: 3d furnished rooms with layouts and semantics},
  author={Fu, Huan and Cai, Bowen and Gao, Lin and Zhang, Ling-Xiao and Wang, Jiaming and Li, Cao and Zeng, Qixun and Sun, Chengyue and Jia, Rongfei and Zhao, Binqiang and others},
  booktitle={Proceedings of the IEEE/CVF International Conference on Computer Vision},
  pages={10933--10942},
  year={2021}
}

@article{yang2025optiscene,
  title={Optiscene: Llm-driven indoor scene layout generation via scaled human-aligned data synthesis and multi-stage preference optimization},
  author={Yang, Yixuan and Luo, Zhen and Ding, Tongsheng and Lu, Junru and Gao, Mingqi and Yang, Jinyu and Sanchez, Victor and Zheng, Feng},
  journal={arXiv preprint arXiv:2506.07570},
  year={2025}
}

@article{mao2025spatiallm,
  title={Spatiallm: Training large language models for structured indoor modeling},
  author={Mao, Yongsen and Zhong, Junhao and Fang, Chuan and Zheng, Jia and Tang, Rui and Zhu, Hao and Tan, Ping and Zhou, Zihan},
  journal={arXiv preprint arXiv:2506.07491},
  year={2025}
}

@article{luo2025empirical,
  title={An empirical study of catastrophic forgetting in large language models during continual fine-tuning},
  author={Luo, Yun and Yang, Zhen and Meng, Fandong and Li, Yafu and Zhou, Jie and Zhang, Yue},
  journal={IEEE Transactions on Audio, Speech and Language Processing},
  year={2025},
  publisher={IEEE}
}

@article{zhu2024model,
  title={Model tailor: Mitigating catastrophic forgetting in multi-modal large language models},
  author={Zhu, Didi and Sun, Zhongyi and Li, Zexi and Shen, Tao and Yan, Ke and Ding, Shouhong and Kuang, Kun and Wu, Chao},
  journal={arXiv preprint arXiv:2402.12048},
  year={2024}
}

@article{bai2024hallucination,
  title={Hallucination of multimodal large language models: A survey},
  author={Bai, Zechen and Wang, Pichao and Xiao, Tianjun and He, Tong and Han, Zongbo and Zhang, Zheng and Shou, Mike Zheng},
  journal={arXiv preprint arXiv:2404.18930},
  year={2024}
}

@inproceedings{jiang2024hallucination,
  title={Hallucination augmented contrastive learning for multimodal large language model},
  author={Jiang, Chaoya and Xu, Haiyang and Dong, Mengfan and Chen, Jiaxing and Ye, Wei and Yan, Ming and Ye, Qinghao and Zhang, Ji and Huang, Fei and Zhang, Shikun},
  booktitle={Proceedings of the IEEE/CVF Conference on Computer Vision and Pattern Recognition},
  pages={27036--27046},
  year={2024}
}

@article{hu2022lora,
  title={Lora: Low-rank adaptation of large language models.},
  author={Hu, Edward J and Shen, Yelong and Wallis, Phillip and Allen-Zhu, Zeyuan and Li, Yuanzhi and Wang, Shean and Wang, Liang and Chen, Weizhu and others},
  journal={Iclr},
  volume={1},
  number={2},
  pages={3},
  year={2022}
}

@article{zhang2025open3d,
  title={Open3D-VQA: A Benchmark for Comprehensive Spatial Reasoning with Multimodal Large Language Model in Open Space},
  author={Zhang, Weichen and Zhou, Zile and Zeng, Xin and Liu, Xuchen and Fang, Jianjie and Gao, Chen and Li, Yong and Cui, Jinqiang and Chen, Xinlei and Zhang, Xiao-Ping},
  journal={arXiv preprint arXiv:2503.11094},
  year={2025}
}

@article{zhou2025il3d,
  title={Il3d: A large-scale indoor layout dataset for llm-driven 3d scene generation},
  author={Zhou, Wenxu and Nie, Kaixuan and Du, Hang and Yin, Dong and Huang, Wei and Guo, Siqiang and Zhang, Xiaobo and Hu, Pengbo},
  journal={arXiv preprint arXiv:2510.12095},
  year={2025}
}

@inproceedings{liu2024mmbench,
  title={Mmbench: Is your multi-modal model an all-around player?},
  author={Liu, Yuan and Duan, Haodong and Zhang, Yuanhan and Li, Bo and Zhang, Songyang and Zhao, Wangbo and Yuan, Yike and Wang, Jiaqi and He, Conghui and Liu, Ziwei and others},
  booktitle={European conference on computer vision},
  pages={216--233},
  year={2024},
  organization={Springer}
}

@article{torres2017fault,
  title={Fault and error tolerance in neural networks: A review},
  author={Torres-Huitzil, Cesar and Girau, Bernard},
  journal={IEEE access},
  volume={5},
  pages={17322--17341},
  year={2017},
  publisher={IEEE}
}

@article{wang2024root,
  title={Root: Vlm based system for indoor scene understanding and beyond},
  author={Wang, Yonghui and Chen, Shi-Yong and Zhou, Zhenxing and Li, Siyi and Li, Haoran and Zhou, Wengang and Li, Houqiang},
  journal={arXiv preprint arXiv:2411.15714},
  year={2024}
}

@article{voulodimos2018deep,
  title={Deep learning for computer vision: A brief review},
  author={Voulodimos, Athanasios and Doulamis, Nikolaos and Doulamis, Anastasios and Protopapadakis, Eftychios},
  journal={Computational intelligence and neuroscience},
  volume={2018},
  number={1},
  pages={7068349},
  year={2018},
  publisher={Wiley Online Library}
}

@inproceedings{stricker1995similarity,
  title={Similarity of color images},
  author={Stricker, Markus Andreas and Orengo, Markus},
  booktitle={Storage and retrieval for image and video databases III},
  volume={2420},
  pages={381--392},
  year={1995},
  organization={SPiE}
}

@article{wang2024mllm,
  title={Mllm can see? dynamic correction decoding for hallucination mitigation},
  author={Wang, Chenxi and Chen, Xiang and Zhang, Ningyu and Tian, Bozhong and Xu, Haoming and Deng, Shumin and Chen, Huajun},
  journal={arXiv preprint arXiv:2410.11779},
  year={2024}
}

@inproceedings{huang2024visual,
  title={Visual hallucinations of multi-modal large language models},
  author={Huang, Wen and Liu, Hongbin and Guo, Minxin and Gong, Neil},
  booktitle={Findings of the Association for Computational Linguistics: ACL 2024},
  pages={9614--9631},
  year={2024}
}

@inproceedings{fei2024fine,
  title={Fine-grained structural hallucination detection for unified visual comprehension and generation in multimodal LLM},
  author={Fei, Hao and Luo, Meng and Xu, Jundong and Wu, Shengqiong and Ji, Wei and Lee, Mong-Li and Hsu, Wynne},
  booktitle={Proceedings of the 1st ACM Multimedia Workshop on Multi-modal Misinformation Governance in the Era of Foundation Models},
  pages={13--22},
  year={2024}
}

@article{kovesi2000phase,
  title={Phase congruency: A low-level image invariant},
  author={Kovesi, Peter},
  journal={Psychological research},
  volume={64},
  number={2},
  pages={136--148},
  year={2000},
  publisher={Springer}
}

@article{fan2022phase,
  title={Phase congruency order-based local structural feature for SAR and optical image matching},
  author={Fan, Jianwei and Ye, Yuanxin and Liu, Guichi and Li, Jian and Li, Yanling},
  journal={IEEE Geoscience and Remote Sensing Letters},
  volume={19},
  pages={1--5},
  year={2022},
  publisher={IEEE}
}

@article{carlinet2026leveraging,
  title={Leveraging the geodesic distance transform on interval-valued maps to compute the tree of shapes},
  author={Carlinet, Edwin and Esteban, Baptiste and Boutry, Nicolas and Romero-Garcia, Gonzalo},
  year={2026}
}

@inproceedings{dai2017scannet,
  title={Scannet: Richly-annotated 3d reconstructions of indoor scenes},
  author={Dai, Angela and Chang, Angel X and Savva, Manolis and Halber, Maciej and Funkhouser, Thomas and Nie{\ss}ner, Matthias},
  booktitle={Proceedings of the IEEE conference on computer vision and pattern recognition},
  pages={5828--5839},
  year={2017}
}

@article{wang2023amsa,
  title={AMSA: Adaptive multimodal learning for sentiment analysis},
  author={Wang, Jingyao and Mou, Luntian and Ma, Lei and Huang, Tiejun and Gao, Wen},
  journal={ACM Transactions on Multimedia Computing, Communications and Applications},
  volume={19},
  number={3s},
  pages={1--21},
  year={2023},
  publisher={ACM New York, NY}
}

@inproceedings{chen2024spatialvlm,
  title={Spatialvlm: Endowing vision-language models with spatial reasoning capabilities},
  author={Chen, Boyuan and Xu, Zhuo and Kirmani, Sean and Ichter, Brain and Sadigh, Dorsa and Guibas, Leonidas and Xia, Fei},
  booktitle={Proceedings of the IEEE/CVF Conference on Computer Vision and Pattern Recognition},
  pages={14455--14465},
  year={2024}
}

@article{hong20233d,
  title={3d-llm: Injecting the 3d world into large language models},
  author={Hong, Yining and Zhen, Haoyu and Chen, Peihao and Zheng, Shuhong and Du, Yilun and Chen, Zhenfang and Gan, Chuang},
  journal={Advances in Neural Information Processing Systems},
  volume={36},
  pages={20482--20494},
  year={2023}
}

@inproceedings{farshad2023scenegenie,
  title={Scenegenie: Scene graph guided diffusion models for image synthesis},
  author={Farshad, Azade and Yeganeh, Yousef and Chi, Yu and Shen, Chengzhi and Ommer, B{\"o}jrn and Navab, Nassir},
  booktitle={Proceedings of the IEEE/CVF International Conference on Computer Vision},
  pages={88--98},
  year={2023}
}

\end{document}